\documentclass[twocolumn]{aastex63}

\usepackage{amsmath,bm}
\usepackage{soul}

\newcommand{\dHybridR}{{\tt dHybridR}}
\newcommand{\del}{{\bf \nabla}}

\newcommand{\Emax}{E_{\rm max}}
\newcommand{\ncr}{n_{\rm cr}}
\newcommand{\vcr}{v_{\rm cr}}
\newcommand{\Jcr}{J_{\rm cr}}
\newcommand{\ecr}{\varepsilon_{\rm cr}}
\newcommand{\mv}[1]{{\bf #1}}

\newcommand{\remove}[1]{}

\graphicspath{{./}{Figures/}}

\shorttitle{\dHybridR}
\shortauthors{Haggerty and Caprioli}

\graphicspath{{./}{figures/}}

\submitjournal{ApJ}

\begin{document}

\title{\emph{dHybridR}: a Hybrid--Particle-in-Cell Code Including Relativistic Ion Dynamics}

\correspondingauthor{Colby C. Haggerty}
\email{chaggerty@uchicago.edu}

\author[0000-0002-2160-7288]{Colby C. Haggerty}
\author[0000-0003-0939-8775]{Damiano Caprioli}
\affil{Department of Astronomy and Astrophysics, University of Chicago, 5640 S Ellis Ave, Chicago, IL 60637, USA}

\begin{abstract}
We present the first plasma simulations obtained with the code \dHybridR{}, a hybrid particle-in-cell code with fluid electrons and both thermal and energetic ions that retain relativistic dynamics. 
\dHybridR{} is constructed to study astrophysical and space-physics problems where a few energetic non-thermal particles (i.e., cosmic rays, CRs) affect the overall dynamics of a non-relativistic plasma, such as CR-driven instabilities, collisionless shocks, magnetic reconnection, turbulence, etc.
In this method paper we provide some applications to linear (resonant/non-resonant CR streaming instability) and strongly non-linear (parallel shocks) problems that show the capabilities of the code. 
In particular, we provide the first self-consistent hybrid runs that show the acceleration of relativistic ions at non-relativistic shocks; CRs develop a power-law in momentum, which translates to a broken power law in energy that exhibits a steepening around the ion rest mass, as predicted by the theory of diffusive shock acceleration.
We present examples of 2D \dHybridR{} runs relevant for fast shocks in radio supernovae, whose evolution can be followed in real time, and 3D runs of low-Mach-number heliospheric shocks, which can be compared with in-situ spacecraft observations.

\end{abstract}

\section{Introduction} \label{sec:intro}
Understanding the generation and dynamical effects of non-thermal, high-energy particles (Cosmic Rays, CRs) in astrophysical plasmas has been an important question since their discovery in the early 20$^{\rm th}$ century \citep[see, e.g.,][for some representative seminal papers on the acceleration of Galactic CRs]{baade+34,Fermi49, chen+75, krymskii77, axford+77, bell78a, bell78b, blandford+78}.
CRs are ubiquitous throughout the universe and in the Galactic interstellar medium are in equipartition with the thermal plasma and the magnetic fields, despite being very few in number, about $10^{-9}$ times less abundant than thermal protons \citep[e.g.,][and references therein]{yoasthull+14}. 

Self-consistent modeling of the non-linear interplay between CRs, thermal plasma, and magnetic fields is a challenging problem and requires kinetic numerical  approaches;
moreover, such a non-linear physics inherently spans multiple length and time scales.
For instance, the gyroradius of a GeV particle is about $10^{12}$cm in the $\mu$G magnetic field typical of heliospheric and interstellar media, significantly larger than electron/ion skin depths, which are of the order of $10^5-10^7$cm for typical densities of about 1 cm$^{-3}$.
Accelerators can be several orders of magnitude larger:  $\sim 10^{9}$cm for the Earth bow shock, a fraction to a few astronomical units for interplanetary shocks, tens of pc for Galactic supernova remnants and even a few Mpc for radio relics in galaxy clusters.

Fully kinetic plasma models (like Particle-In-Cell, hereafter PIC, or Vlasov codes) can accurately model all of the relevant physics in collisionless systems 
by evolving the 6-dimensional phase space distribution function of both ions and electrons \citep[e.g.,][]{bl91,bell+06,valentini+07,lapenta12,palmroth+18}.
However, these fully-kinetic simulations require grid sizes and time steps that resolve both the electron and ion dynamics, and because a electron is a factor of 1836 lighter than a proton, the characteristic scales of the electron dynamics are significantly smaller than the ions'. 
Having to resolve the electron scales limits the ability of such approaches to model the long-term evolution of the ions and, especially, of the CRs.

The hybrid model, which treats ions as kinetic macro-particles that satisfy the Vlasov equation with phase space trajectories evolved by the Lorentz force equation and electrons as a fluid that keeps the system charge neutral, can  bridge thermal and non-thermal regimes at the expense of the detailed kinetic electron physics.
Hybrid models \citep[see][for reviews]{winske+96,Lipatov02} have been used to study many different plasma problems including shocks \citep[e.g.,][]{winske85,quest88,burgess89, giacalone+92,giacalone04,gargate+12,burgess+13,burgess+16,caprioli15p}, turbulence \citep[e.g.,][]{ karimabadi+14, matthaeus+15, pecora+18, arzamasskiy+19} and magnetic reconnection \citep[e.g.,][]{mandt+94, shay+01, le+09}.

An implicit assumption of the hybrid model, however, is that the speed of light is taken to be infinitely large, in order to neglect Maxwell's correction in the Amp\`ere law (see Section~\ref{sec:hybrid} for more details), which forces the ion dynamics to be non-relativistic.
This restriction is significant for modeling the physics of CRs and may raise concerns when simulations are compared with observations. 

Alternative approaches have used a kinetic description of CRs, while treating the thermal population as a magneto-hydrodynamical (MHD) fluid \citep[e.g.,][]{zachary+86,lb00, bai+15, vanmarle+18, mignone+18,dubois+19}.
While these MHD-PIC simulations can capture some CR physics, the gap between thermal and energetic particles requires the injection of CRs in the system to be externally prescribed, rather than modeled from first principles. 

In this work we present the first --to our knowledge-- hybrid code that includes relativistic ion dynamics, \dHybridR{}, which is built upon the massively-parallel Newtonian code \emph{dHybrid} \citep{gargate+07}.
In Section~\ref{sec:hybrid} we outline the basics of the code and we argue that the set of systems with both thermal and CR populations can be modeled this way without violating any of the hybrid approximations.
In Section~\ref{sec:instability} we compare \dHybridR{} simulations of CR streaming instabilities with linear theory predictions and show that the physics of CRs and thermal plasma interaction are being correctly modeled.
In Section~\ref{sec:shocks}, we investigate the acceleration of CRs in parallel shocks and the lack thereof in oblique shocks.
Finally, in Section~\ref{sec:3D}, we show a  3D  simulation of an oblique, low Mach number shock with parameters comparable to the Earth's bow shock, which exhibits features consistent with very recent {\it in-situ} observations \citep{johlander+16b, johlander+18}.

\section{Hybrid and \dHybridR}\label{sec:hybrid}
The hybrid model for simulating collisionless plasma physics is fundamentally a Monte-Carlo approach to solving the Vlasov--Maxwell system of equations:
\begin{gather}
      \label{eq:vlas}
    \frac{\partial f_s}{\partial t} + \mv{v}\cdot \del + \frac{q_s}{m_s}(\mv{E} + \frac{\bf v}{c} \times \mv{B})\cdot \del_v f = 0\\ 
    \frac{\partial \mv{B}}{\partial t} = -c \del \times \mv{E}\\
    \frac{\partial \mv{E}}{\partial t} = c \del \times \mv{B} - 4\pi \mv{J}\\\label{eq:amp}
    \del \cdot \mv{B} = 0\\
    \del \cdot \mv{E} = \sum_s q_sn_s\label{eq:gauss}
\end{gather}
where $\mv{E}$ and $\mv{B}$ are the electric and magnetic fields, $f_s(\mv{x}, \mv{v}, t)$ is the phase-space distribution function for a given species $s$ of particles with charge $q_s$ and mass $m_s$,
$n_s \equiv \int f_s d^3v$ is the number density of species $s$ and $\mv{J} \equiv \sum_s q_s n_s\mv{V}_s$ is the total current, where $\mv{V}_s \equiv \int vf_sd^3v/n_s$ is the bulk velocity of each species.
In this work only electron--proton plasmas will be considered, but ions with arbitrary mass and charge can be easily accounted for \citep[e.g.,][]{caprioli+17}.

The motivation of the hybrid model is to simulate kinetic ion dynamics (i.e., Equation \ref{eq:vlas}) on larger length and time scales at the expense of kinetically modeling electron dynamics. 
In practice this is done by assuming that the electron mass is negligibly small compared to the ion mass. 
In this way, electrons are treated as a massless charge-neutralizing fluid that enforces quasi-neutrality in system. 
This corresponds to  $n_i = n_e$ and hence to $\del \cdot \mv{E} = 0$ (Equation \ref{eq:gauss}) and $\mv{J} = en_i(\mv{V_i} - \mv{V_e})$.

The evolution of the ions in time are described by Equation \ref{eq:vlas}. 
In practice this is done by approximating the ion distribution function with a large number of \emph{macro-particles} whose motion in phase space is determined by the Lorentz force. 
For a given set of electromagnetic fields, the macro-particle position and velocity can be advanced in time. 
The updated positions and velocities can be interpolated onto a grid, returns a fluid density and bulk flow;
note that because the electrons are taken to be massless, they do not contribute to the bulk flow. 

The electric field, ${\bf E}$, is determined by multiplying the Vlasov equation for the electrons (Equation \ref{eq:vlas}) by $m_e \mv{v}$ and integrating it over all of velocity space, which yields:
\begin{equation}
    m_e n_e \left(\frac{\partial}{\partial t} - \mv{V}_e \cdot \del\right)\mv{V}_e = -\del P_e -e n_e\mv{E} - \frac{e n_e \mv{V}_e}{c}\times \mv{B}.
\end{equation}
Here we introduced the (isotropic) electron pressure  $P_e$, which encompasses higher order moments of the electron distribution function. 
Reapplying the assumption that $n_i = n_e$ in the limit $m_e\ll m_i$, we derive an effective Ohm's law for the electric field:
\begin{eqnarray}
{\bf E}& = &-\frac{{\bf V}_e}{c} \times {\bf B} - \frac{1}{en}{\bm{\nabla}}P_e\\
& =& -\frac{{\bf V}_i}{c} \times {\bf B} + \frac{{\bf J}}{enc} \times {\bf B} - \frac{1}{en}{\bm{\nabla}}P_e\label{eq:ohms}
\end{eqnarray}

The next assumption required for the hybrid model is to neglect the displacement current in Amp\`ere's law (i.e., the time derivative of the electric field) such that ${\bm \nabla} \times {\bf B} = \frac{4\pi}{c}{\bf J}$;
this usually referred to as the radiation-free limit, or the \emph{Darwin approximation}. 
This assumption is often equated with taking the speed of light to be much larger than any other velocity in the system;
however, we show below how it may hold even when a small number of relativistic particles are present. 
Ultimately omitting this term from the hybrid model neglects the role of light waves.
Finally, the electron pressure is prescribed by an equation of state, often taken as isotropic and polytropic.
This electric field can then be used in Faraday's law to update the magnetic field and thus yielding a closed set of equations describing the evolution of the systems.  

While the behavior of the ions is fully detailed in the hybrid model, the electrons physical description and evolution is more {\it ad hoc}.
This is evident in choosing the most physically appropriate value of $\gamma_{\rm eff}$ for a polytropic electron equation of state, $P_e \propto n^{\gamma_{\rm eff}}$.
It could be argued that the electrons should be adiabatic and so $\gamma_{\rm eff} = 5/3$.
However, if the adiabatic description is used in shocks with a large Mach number, electrons cannot increases their entropy at the shock and the downstream electron pressure may end up being orders of magnitude smaller than the ion pressure.
If one asserts that the electron and ion downstream pressure should be in equipartition, the Rankine-Hugoniot jump condition may be used to calculate what $\gamma_{\rm eff}$ should be \citep[see the appendix of][for more details]{caprioli+18}.
More complicated, anisotropic, prescriptions may be needed when dealing with magnetic reconnection \citep[e.g.,][]{le+09}.
In this work we will use the equipartition equation of state for shock simulations and the adiabatic one for CR streaming simulations.

Along with the disparate length and time scales, plasma systems can also span multiple scales in velocity space, ranging from thermal particles that make up the bulk of the plasma to CRs with kinetic energies orders of magnitude larger than their rest mass. 
An implicit assumption of the Darwin model is that the bulk velocities of the system are small relative to the speed of light, and because of this hybrid codes have traditionally not included relativistic effects for the macro-particle ions. 
However, since this approximation is based on \emph{bulk motions} being small relative to the speed of light, even plasma systems with a non-relativistic background and a small number of relativistic particles (or CRs) can be modeled in this limit. 
This can be seen from a scaling argument of Amp\`ere's law with Maxwell's correction:
\begin{equation}
    {\bm \nabla} \times {\bf B} = \frac{4\pi}{c}{\bf J} + \frac{1}{c}\frac{\partial {\bf E}}{\partial t} 
    \ \to\  \frac{B}{\lambda} \colon \frac{4\pi J}{c} \colon \frac{E}{c\tau}
\end{equation}
where derivatives have been replaced by $\lambda$ and $\tau$, which correspond to characteristic length and time scales of the systems we are interested in studying, and the colon ($\colon$) denotes an order of magnitude comparison. 
We can see from the scaling of Faraday's law that $E/B \sim V/c$, where $V = \lambda/\tau$ is the characteristic velocity of the system. Using this and that $J \sim enV$ we can simplify our scaling equation to
\begin{equation}
    1 \colon \frac{\lambda}{d_i}\frac{V}{v_A} \colon \left ( \frac{V}{c} \right )^2
\end{equation}
where $d_i = c/\omega_{pi} = \sqrt{c^2m_i/4\pi n e^2}$ is the ion inertial length (skin depth)  and $v_A = B/\sqrt{4\pi m_i n}$ is the Alfv\'en speed. 
Neglecting the displacement current is no longer appropriate when the third term becomes comparable to the other two and so we find that this approximation is good as long as
\begin{equation}\label{eq:speed_cond}
    \left(\frac{V}{c}\right)^2 \ll 1,\ {\rm and} \ \frac{Vv_A}{c^2} \ll 1
\end{equation}
where we used 1 for $\lambda/d_i$, which is the strictest value that can be used for hybrid simulations.
The systems that we aim to study are composed of a background ion thermal population with number density $n_{i}$, characteristic velocity $V_{\rm bkg} \ll c$ and a high-energy CR population with $n_{\rm cr} \ll n_{i}$ and $v_{\rm cr} \sim c$.
The composite background + CR populations bulk flow speed can be estimated as 
\begin{equation}\label{eq:char_flow}
\frac{V}{c} = \frac{n_{i}V_{\rm bkg} + n_{\rm cr}c}{c(n_{i} + n_{\rm cr})}
\approx \frac{V_{\rm bkg}}{c} + \frac{n_{\rm cr}}{n_{i}}
\end{equation}
From Equations \ref{eq:speed_cond} and \ref{eq:char_flow}, we find three conditions the systems must meet for this approximation to be valid: 
\begin{itemize}
    \item $V_{\rm bkg}\ll c$, i.e., bulk flows cannot be relativistic;
    \item $n_{\rm cr}\ll n_{n_i}$, i.e., the CR number density must be negligible relative to the gas number density;
    \item $v_A\ll c$, i.e., magnetic field energy density must be much smaller than the rest mass energy density.
\end{itemize}
 The last condition is derived by taking the bulk flow velocity to be Alfv\'enic; 
 note that $B/\sqrt{4\pi m_i n}$ can even exceed $c$, in which case the dispersion relation for an Alfv\'en wave needs to be modified by including the displacement current term, thus violating one of the previously outlined assumptions for hybrid \citep{krall-trivel}. 
These conditions are satisfied for many systems in space and astrophysical plasmas where CR acceleration, transport and scattering are important.

To study these types of problems we have developed \dHybridR, a hybrid simulation code that retains the fully relativistic ion dynamics. 
\dHybridR{} is a generalization of the \emph{dHybrid} code \citep{gargate+07}, where the relativistic Lorentz force is used for the ion macro-particle evolution, i.e.,
\begin{equation}
m_i\frac{d\gamma {\bf v}}{dt} = q{\bf E} + \frac{q{\bf v}}{c}\times {\bf B}
\end{equation}
where $\gamma$ is the Lorentz factor of a given macro-particle and given by $\gamma = 1/\sqrt{1 - (v/c)^2}$.
This is implemented in the code using the well documented relativistic Boris algorithm \citep[see][for details]{bl91}. 

The equations that govern both the electromagnetic fields and the particle dynamics are normalized to arbitrary magnetic field, $B_0$, and number density, $n_0$.
Lengths are scaled to the ion inertial length based on this density, $L_0 \equiv d_{i0} = c/\omega_{pi0}$, and time to the inverse ion gyro-frequency based on this magnetic field, $t_0 \equiv \Omega_{ci0}^{-1} = \frac{cm_i}{eB_0}$. 
Velocities are normalized to the ratio of the length and time normalizations and so a velocity of unity corresponds to the Alfv\'en speed in the reference magnetic field and density, $v_0 \equiv L_0/t_0 = B_0/\sqrt{4\pi m_i n_0}$. 
Electric fields are normalized to $B_0v_0/c$ and temperatures and energies to $m_i v_0^2$. 
Throughout this work  simulations are initialized such that the unshocked/background plasma have a magnetic field, density and ion/electron temperature of unity and so the simulation units are effectively normalized to the background/upstream plasma parameters, i.e., $v_0 = v_A = v_{th}$ and $d_{i0} = d_i = c/\omega_{pi} = r_{g,th}$, the gyroradius of the thermal ions. 
By normalizing the discretized equations in this way, the speed of light only appears as the ratio $c/v_0$ and then only occurs in the Lorentz factor, $\gamma(v) = 1/\sqrt{1 - (v/v_A)^2(v_A/c)^2}$, in the Lorentz force equation. 
The magnetic field is evolved using a two-step Lax-Wendroff scheme that is second-order accurate in space and time \citep{bl91, Hockney}. 
Further details about the non-relativistic implementation of \dHybridR{} are described in \cite{gargate+07}.

The remainder of this paper is dedicated to the demonstration and validation of \dHybridR{} simulating CR generation and transport for selected plasma systems, in which a small number of highly-energetic ions affects the dynamics. 
We will examine the non-resonant streaming instability (commonly referred to as the Bell instability), the resonant streaming instability, and different regimes of collisionless shocks.

In particular, we will study the transition from non-relativistic to relativistic CR energies in fast non-relativistic shocks;
since the required timestep is inversely proportional to $c/v_A$, we initially focus on shock environments where $v_A$ is rather large, such as radio supernovae, where $V_{\rm sh}\sim 0.1c$, $B_0 \sim 0.1{\rm G}$ and $n_0 \sim 10^3 {\rm cm}^{-1}$ at the peak of the synchrotron emission  \citep[e.g.,][]{cf06}.
These parameters correspond to Alfv\'enic mach numbers $M_A\equiv V_{\rm sh}/v_A\sim  10$ and $c/v_A \sim 100$.
Then, we show simulations of lower-Mach number shocks which are more applicable to heliospheric systems, such as planetary bow shocks and interplanetary shocks triggered by coronal mass ejections, where plasma speeds vary between several hundreds to thousands of km s$^{-1}$,  corresponding to Mach numbers ranging from 1 to 10 and $c/v_A \gtrsim 10^4$ \citep{sheeley+85, cane+03}. 
Despite the limited spatial extent of such heliospheric systems, trans-relativistic and even relativistic particles can be produced in such environments, too \citep[e.g.,][]{reames99, tylka+05, wilson+16, reames13, desai+16a}.

There are numerous astrophysical systems where $c/v_A$ and $c/v_{\rm sh}$ are considerably larger than the simulations presented in this work; however, as long as there is a clear separation of scales between the thermal/Alfv\'enic speed, the speed of the shock, and the speed of light, the underlying physics can be studied fruitfully.
This idea implies that the physical results from these simulations, and \dHybridR{} in general, are potentially applicable to many different astrophysical systems.

\section{Resonant and Non-Resonant Streaming Instability}\label{sec:instability}
To verify that \dHybridR{} correctly simulates the physics relevant to systems with CRs, we present two simulations of the CR-driven streaming instability. 
This occurs when a population of low density energetic CRs drift relative to a thermal population, driving the amplification of magnetic fluctuations perpendicular to the mean field.
The characteristics of the instability are controlled by the CR current density: 
in the weak current limit, CRs trigger the growth of modes that are gyro-resonant with themselves \citep[resonant streaming instability; e.g.,][]{kulsrud+69, skilling75a,bell78a,zweibel03}.
In the strong current limit, instead, the return current in background electrons that is needed to enforce charge neutrality drives modes with wavelengths shorter than the CR gyro-radius \citep[non-resonant or Bell instability; e.g.,][]{bell04, weidl+19a}.

The kinetic theory of these instabilities and the transition between the two has been detailed, e.g., by \cite{amato+09} for a CR distribution $\propto p^{-4}$ in momentum.
In this case the boundary between the two regimes is defined by the parameter \citep[see][]{amato+09}:
\begin{equation}
\label{eq:ineq}
    \Bar{\sigma} \equiv \frac{4\pi}{c}\frac{r_L}{B}J_{\rm cr} = \frac{n_{\rm cr}}{n_i}\frac{p_{\rm min}v_d}{m_i v_A^2},
\end{equation}
where $r_L$ is the gyro-radius of the particles with the minimum momentum in the CR distribution, $p_{\rm min}$, and $J_{\rm cr} = e n_{\rm cr} v_d$ is the CR current, defined by the CR number density $n_{\rm cr}$ and their drift velocity  $v_d$.
For $\Bar{\sigma} \gg 1$ the non-resonant mode grows faster than the resonant one, while for $\Bar{\sigma} \ll 1$ they grow at the same rate \citep{bell04, amato+09}.
In the resonant regime, because CRs have a velocity spread much larger than the drift velocity, both right- and left-handed magnetic fluctuations are driven, while in the non-resonant case only electron-driven right-handed modes are amplified.

We set up \dHybridR{} simulations of the CR streaming instabilities with different $n_{\rm CR}\propto \Bar{\sigma}$ and test both the strong and the weak current regimes. This allows us to probe the non-trivial coupling between CRs, magnetic fields, and thermal background plasma both in a MHD-like (non-resonant) and a purely kinetic (resonant) scenario.

We consider two simulations in periodic domains of size $[L_x,L_y]=[10^4, 5] d_i$ with a uniform magnetic field ${\bf B}=B_0{\bf x}$ and a stationary background population of protons with thermal speed equal to $v_A$.
Superimposed on the background population is a lower-density CR population with a power-law distribution in momentum space $f(p) \propto p^{-4}$ extending from $p_{\rm min}/m_i c = 1$ to $p_{\rm max}/m_i c = 10^4$, which is isotropic in a frame moving with a drift velocity $v_d = 10 v_A$. 
The box transverse size makes the simulations effectively 2D for the thermal background, i.e., it is larger than the gyroradius of thermal ions, but actually 1D in terms of the CR length scales.
In both simulations, the speed of light is set to be $c = 100 v_A$ and there are two grid cells per $d_i$; with 225 and 100 macro-particles per cell used for the background and CR populations, respectively. 
The CR number density relative to the background population is adjusted to trigger either the non-resonant  ($n_{\rm cr}/n_i = 10^{-2}$) or the resonant ($n_{\rm cr}/n_i = 10^{-4}$) instability \citep{bell04, amato+09}.
The time step is chosen to be $dt = 2.5\times 10^{-3} \Omega_{ci}^{-1}$ based on the initial magnetic field such that CRs with $\gamma \gg 1$ and $v \approx c$ do not move more than 1 grid space during each time step.

Each simulation is initialized with a mean magnetic field and no electric fields.
However, because of numerical noise inherent to the finite sampling of the ion distribution, there are initially density and bulk flow fluctuations.
These fluctuations generate electric fields through Ohm's law (Equation \ref{eq:ohms}), which produces perpendicular magnetic perturbations that act as seeds for the unstable modes. 
The amplitude of this noise is controlled by the number of macro-particles per cell and for the simulations presented in this work the noise floor is on the order of $\left < B_\perp^2\right >_{\rm noise} \sim 10^{-4} B_0^2$. 
Changing the number of particles per cell alters the initial noise and changes the time that it takes to achieve saturation, but does not affect either the wavelength or the growth rate of the fastest growing modes.

For the non-resonant (or Bell) regime, the fastest growing mode is right handed (hereafter $k_{\rm max}^+$) and its corresponding growth rate, $\gamma^+_{\rm max}$, reads \citep{bell04}:
\begin{equation}
    \frac{\gamma^+_{\rm max}}{\Omega_{ci}} =  k^+_{\rm max}d_i =
    \frac{1}{2}\frac{n_{\rm cr}}{n_i}\frac{v_d}{v_A},
    \label{eq:bell}.
\end{equation}
Instead, in the resonant regime the fastest growing modes have no preferential helicity and their wavenumbers and growth rate read:
\begin{equation}\label{eq:res}
        k^{\pm}_{\rm max}d_i = \frac{m_iv_A}{p_0};\qquad
    \frac{\gamma^\pm_{\rm max}}{\Omega_{ci}} \approx 
    \frac{\pi}{8}\frac{n_{\rm cr}}{n_i}\frac{v_d}{v_A},
\end{equation}
where the $\pm$ superscripts refer to the right and left handed modes, respectively;
Equation \ref{eq:res} is calculated by Taylor-expanding equation 28 in \cite{amato+09} in terms of the small parameter $n_{\rm cr}v_d p_0/(n_im_iv_A^2)$ and keeping only the linear term.

\begin{figure}[t]
\includegraphics[trim=0px 0px 0px 0px, clip, width=0.48\textwidth]{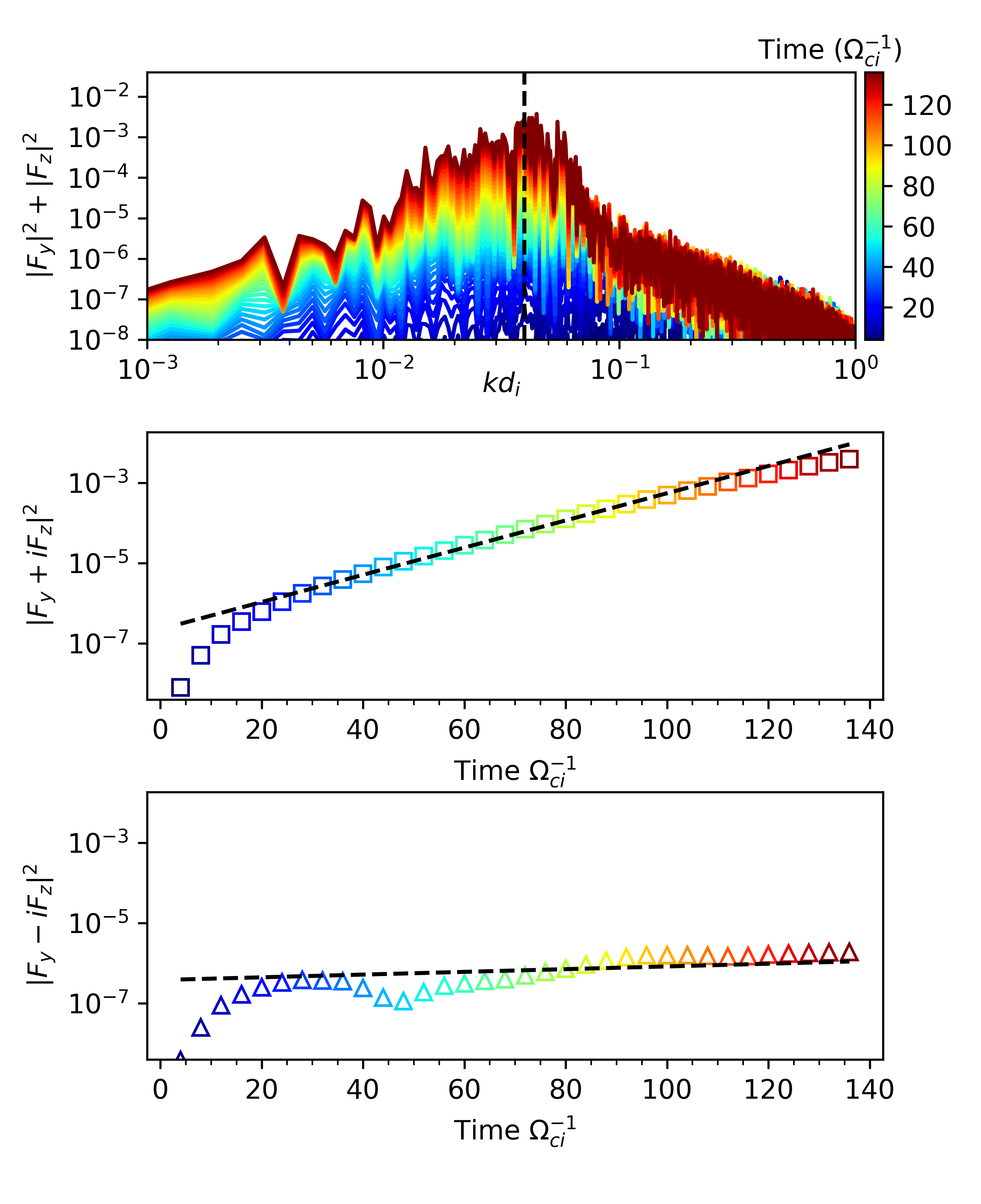}
\caption{Perpendicular magnetic energy spectrum, $|F_y|^2 + |F_z|^2$}, as a function of wave number $k$ and time for a 1D simulation of the non-resonant streaming instability.
Top panel: spectrum as a function of $k d_i$, where each color corresponds to a different time in the simulation; 
the vertical black dashed line corresponds to the $k_{\rm max}$ predicted by the linear theory (Equation \ref{eq:bell}). 
Middle and bottom panels: Magnetic power in both right-handed ($|F_+|^2 = |F_x + iF_z|^2$) and left-handed ($|F_-|^2 =|F_x - iF_z|^2$)  modes as a function of time;
the dashed lines show the growth rates predicted by the linear theory \citep[equation 28 of][]{amato+09}.\label{fig:bell}
\end{figure}

\begin{figure}[t]
\includegraphics[trim=0px 0px 0px 0px, clip, width=0.48\textwidth]{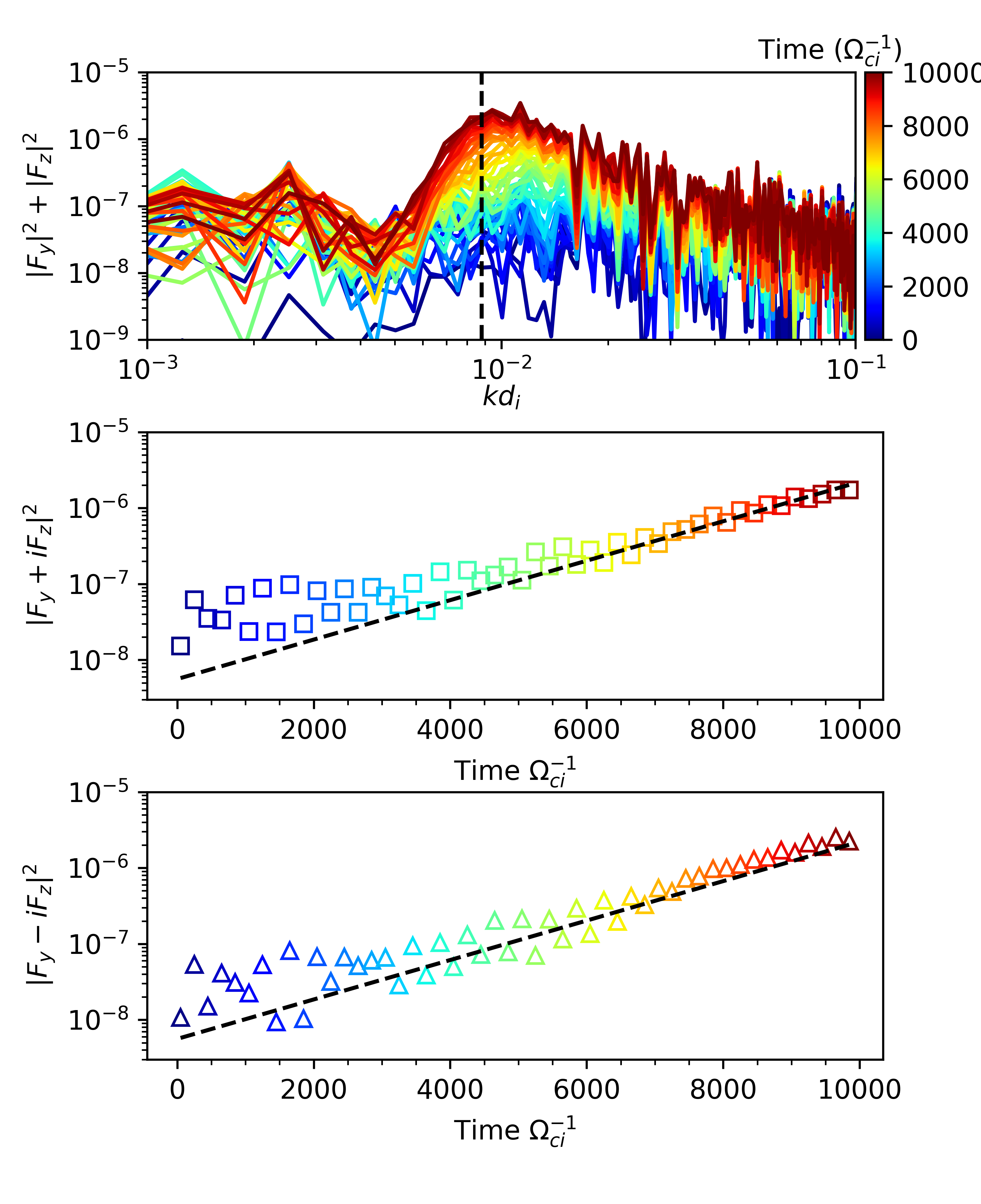}
\caption{As in Figure \ref{fig:bell} for a 1D simulation of the resonant streaming instability. 
The theoretical expectations are from Equation \ref{eq:res}.}\label{fig:ressonant}
\end{figure}

To compare these predictions with the simulations, we introduce 
$F_{i} = {\rm FFT}[B_{i}]$, for $i=y,z$, where ${\rm FFT}$ is the discreet fast Fourier transform calculated along the $x$ direction.
The magnetic power spectrum $|F_y|^2+|F_z|^2$ is plotted in the first panels of Figure~\ref{fig:bell}  and  Figure~\ref{fig:ressonant} for the non-resonant  and resonant cases, respectively.
In both figures the color corresponds to different times in the simulation and the black dashed line shows $k_{\rm max}$ predicted by Eq.\ref{eq:bell} and \ref{eq:res}. There is good agreement between theory and simulation for the location of the fastest growing modes.

The second and third panels of Figure~\ref{fig:bell} and Figure~\ref{fig:ressonant} show the value of the magnetic power in right ($|F_+|^2 = |F_y + iF_z|^2$) and left ($|F_-|^2 = |F_y - iF_z|^2$) handed modes as a function of time for the value of $k_{\rm max}$ denoted by the black dashed line in the first panel.
The magnetic energy is expected to increase exponentially in time as $|F_\pm|^2 \propto e^{2\gamma^\pm_{\rm max} t}$ and the black dashed line corresponds to the $2\gamma_{\rm max}$ given by Equation \ref{eq:bell} and \ref{eq:res};
there is a general agreement between theory and simulations in both the non-resonant and resonant cases. 
Note that the black dashed line in the bottom panel of Figure \ref{fig:bell} is calculated using Equation 28 in \cite{amato+09}.

It is worth noting the differences in the time and length scales of the two instabilities simulated.
The resonant instability stems out from a gyro-resonant interaction with the CR population \citep[e.g.,][]{kulsrud+69}
and amplifies magnetic fluctuations on  scales comparable to the CR gyroradius.

Note that, since $\Bar{\sigma}\ll 1$ for the resonant instability, the growth rates are small compared to the cyclotron frequency of the background population
(Equation \ref{eq:res});
yet, \dHybridR{} can accurately capture this phenomenon over more than $10^4$ cyclotron times (about $4\times 10^6$ time steps).

Recent works have tackled the study of  the non-resonant instability with PIC and hybrid simulations \citep[e.g.,][]{ohira+09, riquelme+09, gargate+10} and of the resonant instability with PIC and PIC-MHD simulations \citep[e.g.][]{Bai+19, holcomb+19, weidl+19b}; these studies have generally found results consistent with theory for the fasting growing mode and corresponding growth rate for the linear phase. 
Nevertheless, the saturation of the CR streaming instability is a complex and non-linear physical phenomenon that is not yet completely understood.
A detailed examination of properties of the two CR streaming instabilities using \dHybridR{}, as well as a more thorough comparison with previous works, is in preparation \citep[see][for preliminary results]{Zacharegkas+19, haggerty+19p}.
The agreement between simulations and the linear theory verifies that \dHybridR{} can accurately model the physical coupling of the thermal background plasma and a drifting CR population for quasi-linear problems, both in the strong and week current regimes.

\section{Non-Relativistic Shocks}\label{sec:shocks}
\subsection{Setup and Simulation Parameters}
\begin{figure*}[!t]
\includegraphics[width=.99\textwidth,clip=true,trim= 0 0 0 0]{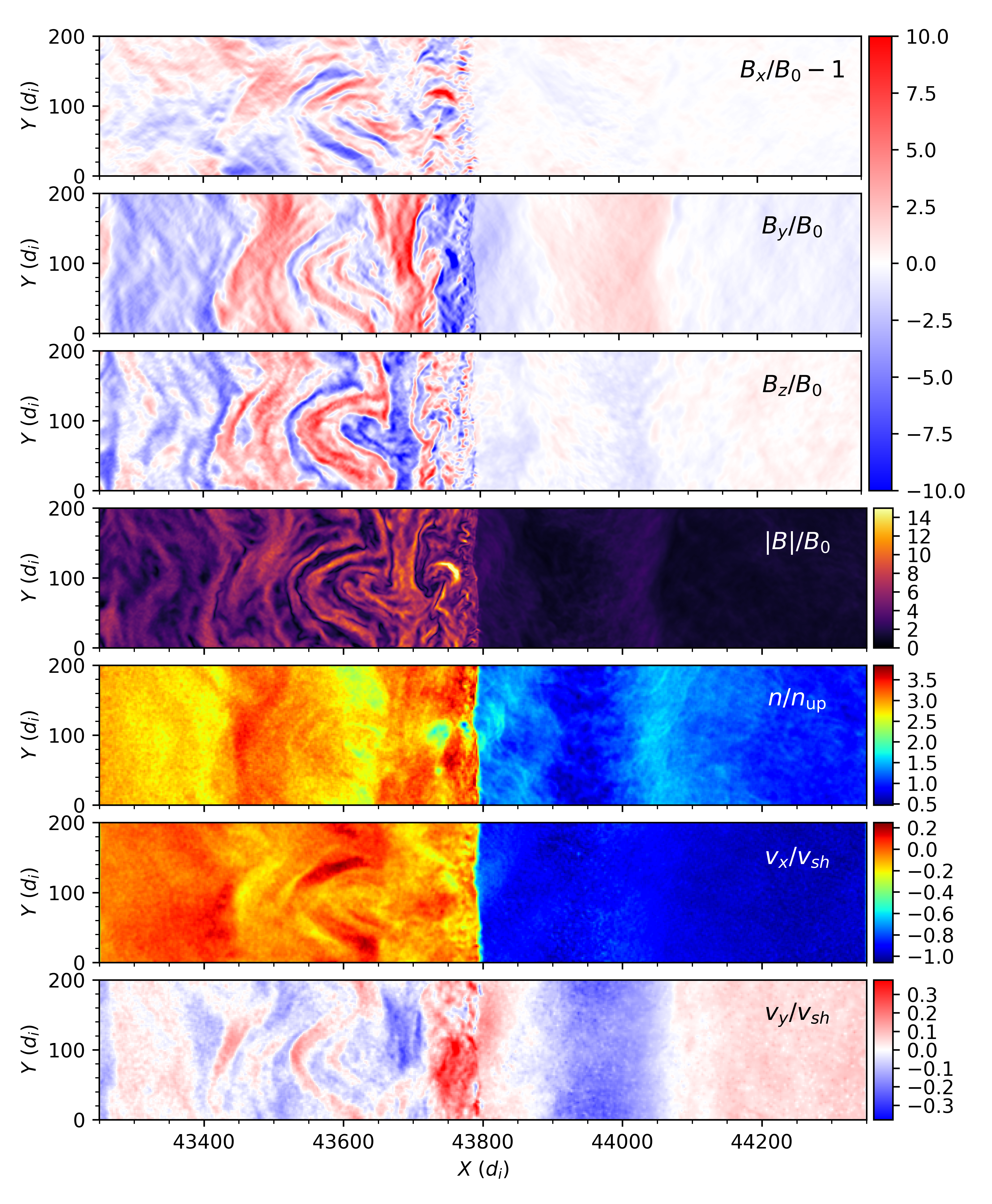}
\caption{2D plasma/fluid quantities around the shock at $t=5560 \Omega_{ci}^{-1}$ from Run A in Table~\ref{tab:shocks}. 
From top to bottom: the 3 components (x,y and z) of the magnetic field in excess to the background (${\bf B} - {\bf B_0}$), magnitude of the magnetic field, density,  normal ($x$) transverse ($y$) bulk flow. All of the quantities are normalized to the upstream values.}\label{fig:overview}
\end{figure*}

Shock simulations were performed with \dHybridR{} following the set up described in \cite{gargate+12}. The simulations are performed in 2.5D (2D in real space, and 3D in momentum space) on a regular Cartesian grid, with periodic boundary conditions in the $y$ direction (transverse to the shock), open on the right boundary ($+x$ direction or normal and upstream of the shock), and a conducting reflecting wall on the left boundary ($-x$ direction and downstream of the shock).
The derivative along x of $E_x,\ B_y,\ {\rm and}\ B_z$ through the left boundary is zero, while $E_y = E_z = 0$ and $B_x = B_x(t=0)$ in the wall.
The shock is formed by initializing the plasma with a bulk flow in the $-x$ direction;
the plasma closest to the left wall is reflected and begins streaming in the $+x$ direction. 
This configuration is unstable and within $\sim10 \Omega_{ci}^{-1}$ a shock forms and travels upstream. 
Across the shock fluid quantities satisfy the Rankine--Hugoniot (RH) jump conditions.
For the simulations in this study, the initial/upstream magnetic field and density are set to unity and the initial magnetic field points in the first quadrant of the $x,y$ plane, the shock angle is measured relative to the positive x direction (normal to the shock) $\vartheta_{Bn}$ (e.g., for a parallel shock ${\bf B} = B_0 {\bf \hat{x}}$ and $\vartheta_{Bn} = 0$). 
The initial ion thermal velocity is equal to the upstream Alfv\'en speed and the electron temperature is equal to the ion temperature ($T_0 = T_i = T_e$). 
Following previous hybrid shock simulations \citep[e.g.,][]{gargate+12, caprioli+14a}, a polytropic index for the electron equation of state is selected so that the downstream electron thermal energy will be half of the upstream kinetic energy in the shock frame \citep[also see][for more details]{caprioli+18}. 
\begin{table}[t]
    \centering
    \begin{tabular}{|c|c|c|c|c|c|c|c|c|}
        \hline
        Run & $M$ & $c/V_A$ & $L_x/d_i$ & $L_\perp/d_i$ & $\Delta x/d_i$ & $\Delta t \Omega_{ci}$ & $\vartheta_{Bn}^{\circ}$\\
        \hline
        \hline
        A & 20 & 200 & $8\times 10^5$ & 200  & 0.5 & .0025 & 0\\
        \hline
        B & 15 & 50 & $10^5$ & 150  & 0.5 & .005 & 0 \\
        \hline
        C & 30 & 10000 & $10^4$ & 2700  & 0.5 & .0025 & 70 \\
        \hline
        3D & 5 & 100 & 1000 & 100 & 0.5 & .02 & 70 \\
        \hline
    \end{tabular}
    \caption{Parameters for the shock simulations presented in this work. From left to right: Alfv\'enic mach number (i.e., $v_{\rm sh}/V_A$), speed of light, longitudinal ($L_x$) and transverse ($L_\perp$) box sizes, spacial grid resolution, time step and angle of the initial magnetic field relative to the upstream plasma bulk flow.
    Note, the time step in simulations C and 3D are set by the speed of fastest particles in the simulation, not the speed of light.}
    \label{tab:shocks}
\end{table}

Shocks are parametrized by their Alfv\'enic and sonic Mach numbers, $M_A = v_{\rm sh}/v_A$ and $M_s = v_{\rm sh}/v_s = v_{\rm sh}/\sqrt{2\gamma k_B T_0/m_i}$, where $v_{\rm sh}$ is the upstream velocity in the lab/simulation frame (i.e., in the frame where the downstream medium is at rest).
The choice of temperature in these simulations links the two Mach numbers, $M_A = \sqrt{10/3}M_s$ and in this work we will reference the Mach number as simply $M \equiv M_A \simeq M_s$. 
We use $2$ grid cells per $d_{i}$ and 4 particles per cell. 
The time step is chosen such that the fastest ion will travel at most one grid cell in one time step. For the parallel shock simulations this corresponds to $\Delta x/ \Delta t < p_{\rm max}/\gamma_{\rm max} \lesssim c$ and $\Delta x/ \Delta t < 3v_{sh}$ for the perpendicular case.

Simulations were run for thousands of cyclotron times to model the CRs transition from non-relativistic to relativistic energies. 
The largest and longest run of these simulations is shown in Figure~\ref{fig:overview} at the end of the simulation, which shows various plasma/fluid quantities around the shock. 
For this run we used $M = 20$, $c = 200 V_{A0}$ and $[L_x,L_y] = [8\times 10^5, 200] d_{i0}$.
The speed of light limiting the fastest speed in our simulation allowed us to run unprecedentedly-long hybrid simulations of non-relativistic shocks, up to $\sim 6000 \Omega_{ci}^{-1}$ before the highest-energy CRs began to escape from the box.

\subsection{Momentum and Energy}
\begin{figure}[htp]
\includegraphics[width=.48\textwidth,clip=true,trim= 0 0 0 0]{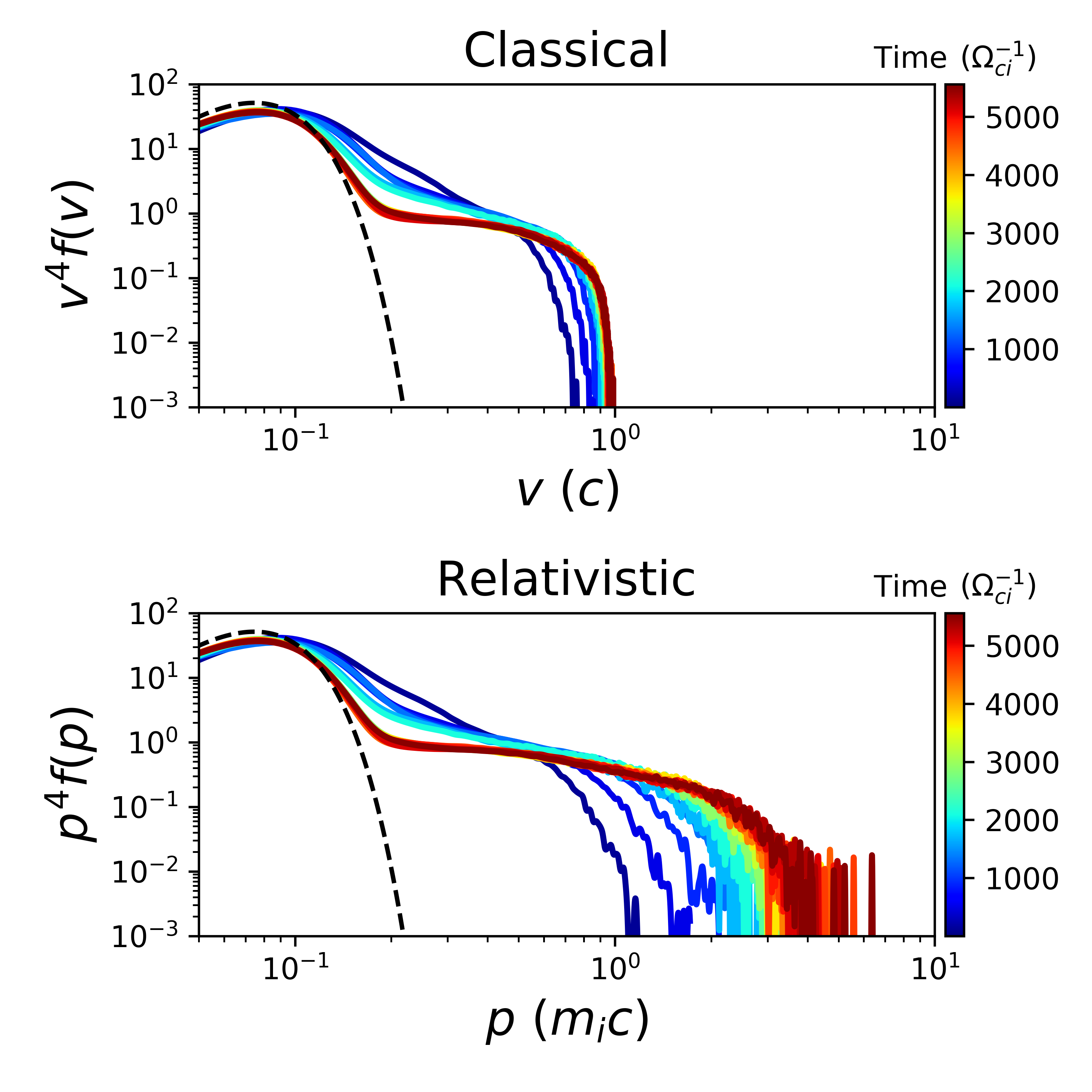}
\caption{Velocity and momentum spectra calculated downstream of the shock (top and bottom panels) for Run A. Different colors correspond to different times in simulation as detailed by the color bars. Velocity and momentum are normalized by $c$ and $m_ic$, and spectra are multiplied by $v^4$ and $p^4$, respectively, for comparison with  the standard DSA prediction. The black dashed line shows a Gaussian with temperature $\sim 20\%$ lower than the temperature predicted by the RH conditions, which compensates for the energy that goes into accelerated ions in the power-law tail.}\label{fig:p}
\end{figure}

Consistent with results from previous non-relativistic hybrid simulations of parallel shocks \citep[e.g.,][]{giacalone+97,burgess+12,caprioli+14a,caprioli+14b,caprioli+14c}, we find that thermal ions can be spontaneously energized into an extended power-law distribution. 
Figure~\ref{fig:p} shows the post-shock distribution function as a function of both the ion velocity normalized to $c$ (first panel) and the ion momentum normalized $m_ic$ (the second panel). 
The majority of ions are thermally heated by the shock, forming the Gaussian peak around $p/m_i \sim v \sim 0.1c$;
the black dashed lines correspond to a Gaussian with temperature reduced by  $20\%$ with respect to the one predicted by the RH conditions for a mono-atomic ideal gas.
The deviation form the prediction is consistent with the amount of energy (about $10-20\%$ of the shock ram energy) that is channeled in the non-thermal power-law distribution that develops beyond $v\sim 0.2c$, whose extent increases with time (color code).
The velocity spectra cuts off at $v \leq c$ as expected, however the momentum continues to extend with the same slope beyond $p\gtrsim m_i c$. 
The distributions shown in Figure~\ref{fig:p} are multiplied by $v^{-4}$ and $p^{-4}$, respectively.
The very reason why the spectrum looks a bit steeper than $p^{-4}$ has profound physical reasons, which will be discussed in a forthcoming paper \citep[see][for a preliminary discussions]{caprioli+19p}.
 
While the momentum spectra shows a nearly constant power law slope, the energy spectrum should have different slopes in the non-relativistic and relativistic regimes. 
The energy distribution is linked to the momentum distribution through the conservation of the phase-space volume: $f(E) = 4\pi p^2 f(p) dp/dE$. In the non-relativistic regime, $E \propto p^2$ and so for a momentum power-law index of $q$, the energy distribution should go as $f(E) \propto E^{(1-q)/2}$, i.e., $E^{-1.5}$ for $q = 4$. 
In the relativistic regime, $E \propto p$ and thus the kinetic energy distribution should scale as $f(E) \propto E^{2-q}$, i.e., the canonical $E^{-2}$ for $q = 4$.

\begin{figure}[htp]
\includegraphics[width=.49\textwidth,clip=true,trim= 0 0 0 0]{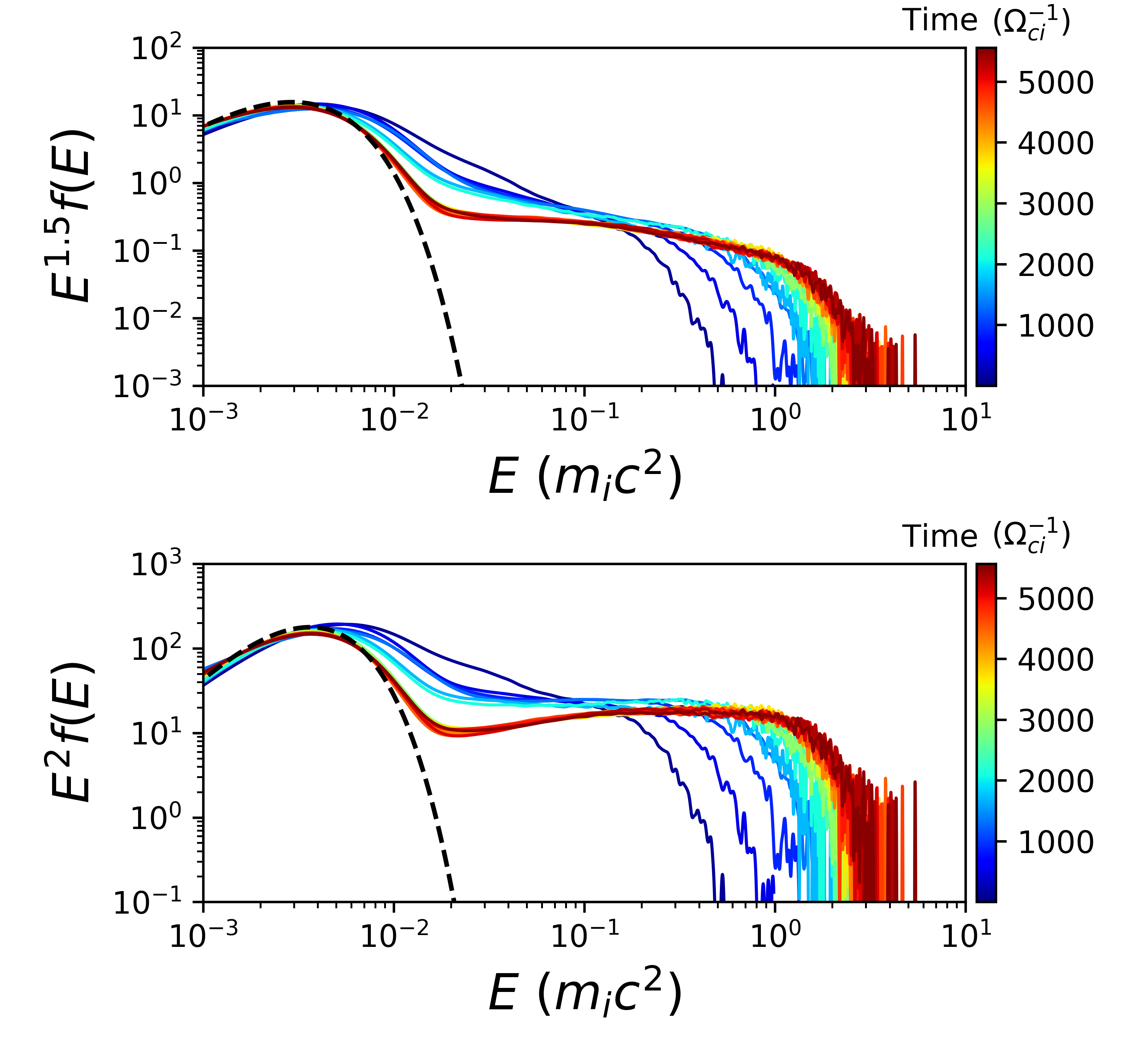}
\caption{Post-shock spectra as a function of the kinetic energy spectrum normalized to $m_ic^2$ (i.e., $\gamma -1$) for Run A in time. 
Distributions are multiplied by $E^q$, where q is the expected power law in the non-relativistic ($E^{1.5}$) and relativistic ($E^2$) regimes (top and bottom panels, respectively).}\label{fig:E}
\end{figure}
\begin{figure}[htp]
\includegraphics[width=.46\textwidth,clip=true,trim= 0 0 0 0]{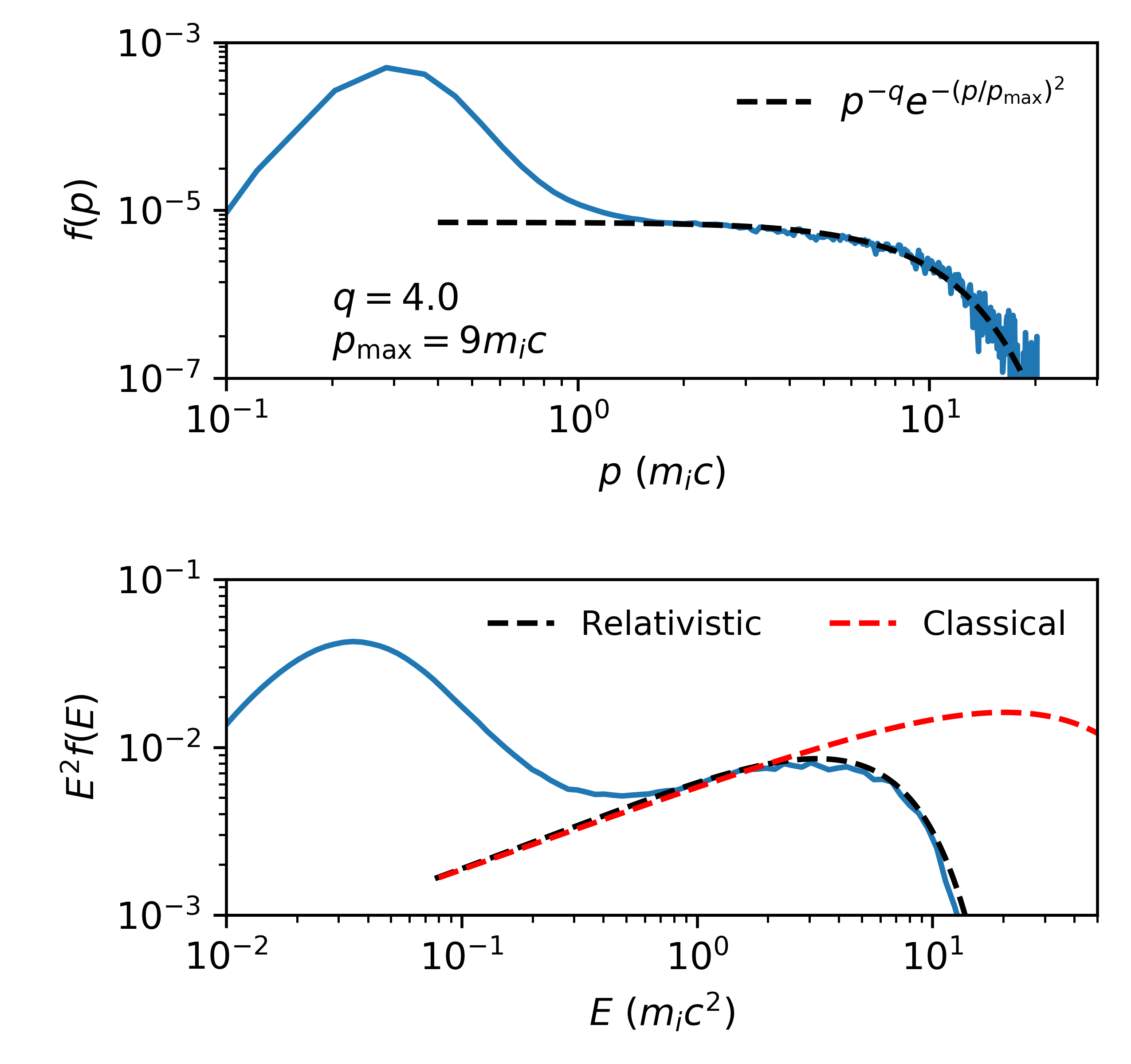}
\caption{Post-shock momentum and energy distributions in Run B at $t=2000 \Omega_{ci}^{-1}$ (top and bottom panels, respectively). The high-momentum tail is well fitted by a distribution $f(p)\propto p^{-q}\exp(-p/p_{\rm max})^2$ with $q=4$ and $p_{\rm max} = 8 m_i c$. 
Such a fitting curve in converted to an energy fitting using either the classical ($E = p^2/2m_i$, red line) or the relativistic ($E = m_i c (\sqrt{p^2/c^2 + 1} -1)$, black line) scalings. The energy distribution is multiplied by $E^2$ to emphasize the transition from $E^{-1.5}$ to $E^{-2}$.}\label{fig:Ecutoff}
\end{figure}

The energy spectrum for our benchmark simulation is shown in Figure~\ref{fig:E}, where the spectrum is multiplied by $E^{1.5}$ in the top panel and $E^2$ in the bottom panel in an attempt to emphasize the agreement with the expected slopes in both the non-relativistic and the relativistic regimes.
In this run some particles became relativistic, with $\gamma\gtrsim 5$, but running such a large simulation long enough for the power-law tail to extend beyond $\sim 10 m_ic^2$ is computationally impractical. 
Thus, in order to see the transition in the energy power-law slope more clearly, we performed a simulation with a smaller speed of light relative to the shock velocity and Alfv\'en speed (Run B in Table~\ref{tab:shocks});
the reduced separation of scales allows us to investigate the trans-relativistic regime more easily.
Figure~\ref{fig:Ecutoff} shows the momentum and energy distribution for Run B at $t=2000 \Omega_{ci}^{-1}$ (top and bottom panels, respectively).
In the first panel the distribution is fitted with to a power-law $\propto p^{-4}$ multiplied by an exponential cut off at $p_{\rm max} = 9 m_i c$. 
The bottom panel, shows the energy distribution, multiplied by $E^2$; 
the black and red dashed lines correspond to the relativistic and classical power-law predictions based on the fit curve from the top panel. 
In essence, the black line shows the shape of the distribution if $E = m_ic^2(\sqrt{p^2/c^2 + 1} -1)$ and the red line for $E = p^2/2m_i$. 
In the non-relativistic regime, both the black and red predictions agree well with the measured spectrum; 
as the distribution extends into the relativistic regime ($E \gtrsim 2 m_i c$), though, there is a clear steepening to the slope of $-2$, followed by the exponential cutoff, which agrees well with the black line prediction.
The classical prediction (red line), however, continues to increase for nearly an order of magnitude in energy beyond the actual energy cut-off.

This analysis shows, for the first time in hybrid simulations, how non-relativistic shocks can accelerate particles to ultra-relativistic energies (with Lorentz factors up to $\gamma\gtrsim 20$ in our case), also confirming that DSA produces power-laws \emph{in momentum space} across the non-relativistic and relativistic regimes.  
These results are consistent with those obtained for both electrons and ions in 1D full-PIC simulations of non-relativistic shocks \citep{park+15} and for electrons in full-PIC simulations of trans-relativistic shocks \citep{crumley+19}. 

One important astrophysical application that stems put from these preliminary runs is relevant for young SNe.
In fact, if we consider the typical values for the magnetic field inferred in type Ib \& Ic supernova with Wolf-Rayet star progenitors \citep{cf06}, the inverse cyclotron time $\Omega_{ci}^{-1}$ would be on the order of milliseconds. 
Both the A \& B Runs have $M_A$ and $c/v_A$ typical of these systems and so physically these simulations are modeling a few seconds of fast radio SNe. 
Notably, in these simulations, thermal protons are accelerated to multi-GeV energies in a matter of seconds, which has implications for the generation of $\gamma-$rays and neutrinos, as discussed below. 

\subsection{Rate of Maximum Energy Increase} \label{sec:maxE}
An important question regarding DSA is what is the maximum energy, $\Emax(t)$, of the particles produced by a shock with a given speed and magnetic field in a finite amount of time \citep[e.g.,][]{drury83,lagage+83a,blasi+07}.
When the magnetic field perturbations responsible for particle diffusion is self-generated by the CRs, $\Emax$ is determined by the current in CRs streaming in the upstream medium, $\Jcr$;
such a current can be estimated as the number density $\ncr$ of particles close to the instantaneous $\Emax$, times their velocity, $\vcr$. 
For a momentum spectrum $f(p)\propto p^{-4}$, in the non-relativistic regime, one has $\ncr\propto p_{\rm max}^3f(p)\propto \Emax^{-1/2}$ and $\vcr\propto \Emax^{1/2}$, so that $\Jcr = e\ncr\vcr\simeq $ is constant in time.
Conversely, in the relativistic regime $\ncr\propto \Emax^{-1}$ and $\vcr\simeq c$, so that  $\Jcr\propto \Emax^{-1}$;
therefore, the current decreases when the maximum CR energy increases and one may expect a slower amplification of the magnetic field, and in turn a slower rate of increase of $\Emax$.
Note that this effect may be partially compensated by the fact that the CR precursor becomes larger when $\Emax$ increases, so that the time available for growing the field (of the order of one advection time on a CR diffusion length) also increases. 
The net effect in general depends on whether most of the field growth is provided by escaping or diffusing particles, and on the details of the instability saturation \citep{caprioli+14b}.

A change in the rate of increase of $E_{\rm max}$ when ions become relativistic was first investigated by \cite{bai+15} using a MHD-PIC approach.
Note that such a framework requires to specify a priori the fraction of particles that effectively become CRs but ---when acceleration becomes efficient--- this fraction has to decrease with time to avoid an energy runaway.
Since a quantitative theory of how this occurs is still lacking, MHD-PIC methods cannot investigate the long-term evolution of the shock self-consistently.

\begin{figure}[htp]
\includegraphics[width=.49\textwidth,clip=true,trim= 0 0 0 0]{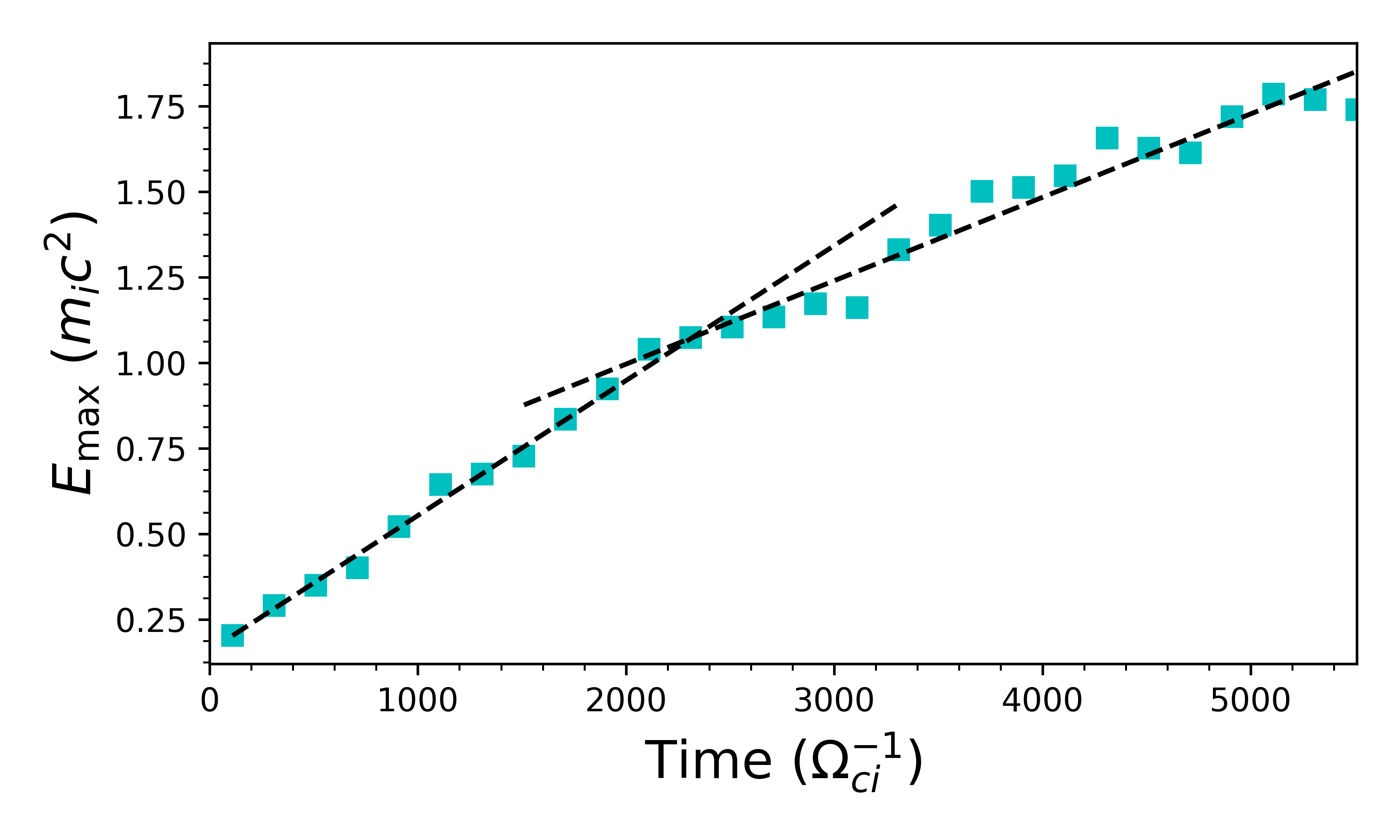}
\caption{Evolution of the maximum CR energy (Equation \ref{eq:Emax}, in units of $m_i c^2$) for Run A in Table~\ref{tab:shocks}.}\label{fig:Emax}
\end{figure}

To quantify the change in the maximum energy we define $\Emax$ as the exponential cutoff of the CR distribution, taken in the form $f(E) \sim E^{-q} e^{-E/\Emax}$. 
Following \cite{bai+15}, we calculate $\Emax$  by integrating over the energy distribution function:
\begin{equation}
\Emax \sim \frac{\int E^{4} f(E) dE}{\int E^{3} f(E) dE}.\label{eq:Emax}
\end{equation}
Since $f(E)$ has an energy slope between $1.5$ and $2$, the integral differs from $\Emax$ by a constant of order unity. 
Figure \ref{fig:Emax} shows the maximum energy as a function of time for Run A, where $\Emax(t)\propto t$ can be fitted with a broken linear function with a change of slope in the trans-relativistic regime. 
The rate of increase of $\Emax$ is about $3.9\times 10^{-4} m_ic^2\Omega_{ci}$ and $2.4\times 10^{-4} m_ic^2\Omega_{ci}$ below and above the rest mass energy, respectively.
This decrease by nearly a factor of two is quantitatively consistent with the reduction found in \cite{bai+15}, further supporting the idea that the decrease is due to a reduction in CR current in the relativistic regime.

The connection between the self-generated diffusion and the growth of the maximum CR energy can be made more explicit by measuring the average diffusion coefficient upstream of the shock. 
For DSA, the return  time upstream is typically the bottleneck of the acceleration rate.
Such a diffusion coefficient $D(E)$ is estimated using the approach outlined in \cite{caprioli+14c}, namely:
\begin{equation}
    D(E) \simeq \frac{v_{\rm sh}}{f_{\rm sh}(E)}\int_{\rm shock}^{x_0} f(x, E) dx \label{eq:diff_coef}
\end{equation}
where $x_0$ is a position far enough upstream that the CR population is negligible and $f_{\rm sh}(E)$ is the CR distribution function just downstream of the shock.

\begin{figure}[htp]
\includegraphics[width=.49\textwidth,clip=true,trim= 0 0 0 0]{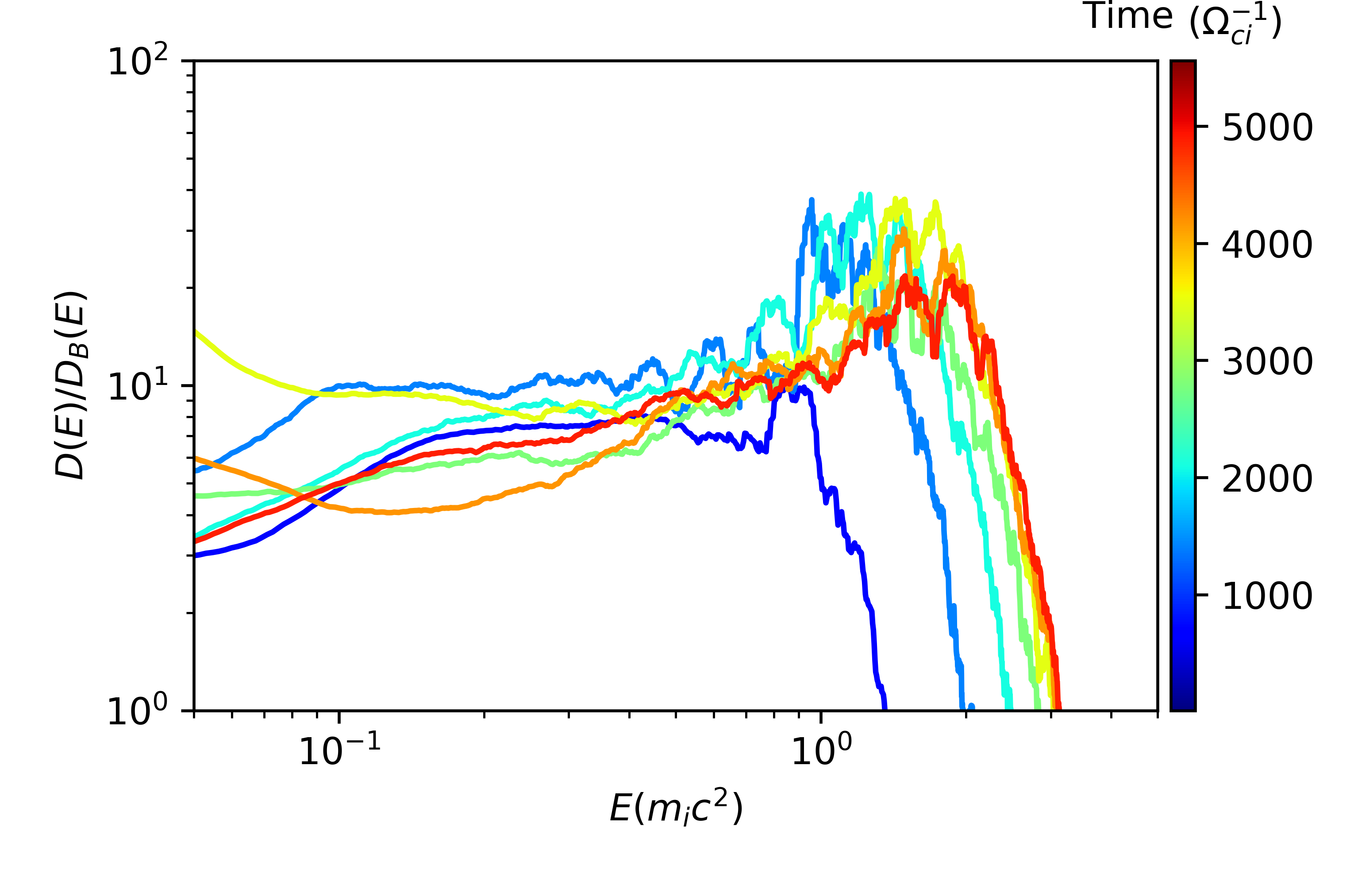}
\caption{Effective diffusion coefficient (Equation \ref{eq:diff_coef}) normalized to the Bohm diffusion coefficient as a function of energy for Run A in Table~\ref{tab:shocks}. The different color lines correspond to different times in the simulation.}\label{fig:diff}
\end{figure}

Figure~\ref{fig:diff} shows the time evolution of the diffusion coefficient normalized to the Bohm diffusion coefficient ($D_B\equiv vr_L/2$) for Run A. 
As discussed in \citep{caprioli+14c}, for $M=20$ the diffusion coefficient is about an order of magnitude larger than Bohm, which is consistent with having self-generated magnetic fluctuations ---at scales resonant with the CRs--- that are approximately an order of magnitude smaller than the initial upstream magnetic field. 
As the simulation evolves in time and the maximum energy transitions into the relativistic regime and we can see a change in the diffusion coefficient. We find the value of the diffusion coefficient is consistently larger at relativistic energies, $D(m_ic^2) \sim 10 - 20 D_B$, than at non-relativistic energies, $D(m_ic^2/5) \sim 5 - 10 D_B$.
The rate of maximum CR  energy increase for non-relativistic shocks can be written as \citep{caprioli+14c}:
\begin{equation}
    \frac{\Emax(t)}{\frac12 m_i v_{\rm sh}^2} \approx \frac13 \frac{D_B(\Emax)}{D(\Emax)}\Omega_{ci} t.
\end{equation}
The increase in the diffusion coefficient as CRs transition to relativistic energies is consistent with the reduction in the rate of change of $\Emax$, as seen in Figure~\ref{fig:Emax}. The increase of the diffusion coefficient by an approximate factor of 2 agrees with the reduction of the slope by a comparable factor.

In previous hybrid simulations, the self-generated diffusion coefficient normalized to the Bohm coefficient has been linked to the Mach number by $D/D_B \propto 1/\sqrt{M}$  \citep{caprioli+14b,caprioli+14c}. 
Using this scaling along with the measured rate of increase of the maximum energy from our simulation, we can calculate a prediction for the maximum energy as a function of time:
\begin{equation}
    \frac{\Emax}{\rm GeV} \approx 20 \left( \beta_{\rm sh}^5\frac{n}{{\rm cm}^{-3}} \frac{B}{{\rm Gauss}} \right)^{1/2} \frac{t}{{\rm s}}
\end{equation}
where $\beta_{sh} \equiv v_{\rm sh}/c$. Again, for the typical values of fast radio supernovae, with $\beta_{\rm sh} \gtrsim 0.01$, CRs with GeV energies will be reached within seconds, and TeV CRs will be produced in about an hour.
If the circumstellar medium is dense enough, multi-TeV neutrinos\footnote{Hadronic neutrinos and $\gamma$-rays of energy $E$ are produced by parent protons of energy $\sim 10E$.} in the range of sensitivity of Ice Cube could be produced in a matter of days after the SN explosion.

\subsection{Acceleration Efficiency}
We consider the evolution of the fraction of shock energy that is transferred to CRs as a function of time.
Following \cite{caprioli+14a} and \cite{caprioli+15}, we distinguish the CRs as the ions that achieved energies $E\gtrsim 10 E_{sh}$ and define the acceleration efficiency $\ecr$ as the fraction of the energy density in these particles normalized by the total energy density.
\begin{equation}
    \varepsilon_{cr} = \frac{\int_{10E_{sh}}^\infty Ef(E)dE}{\int_0^{\infty} Ef(E)dE}
\end{equation}
 
\begin{figure}[htp]
\includegraphics[width=.49\textwidth,clip=true,trim= 0 0 0 0]{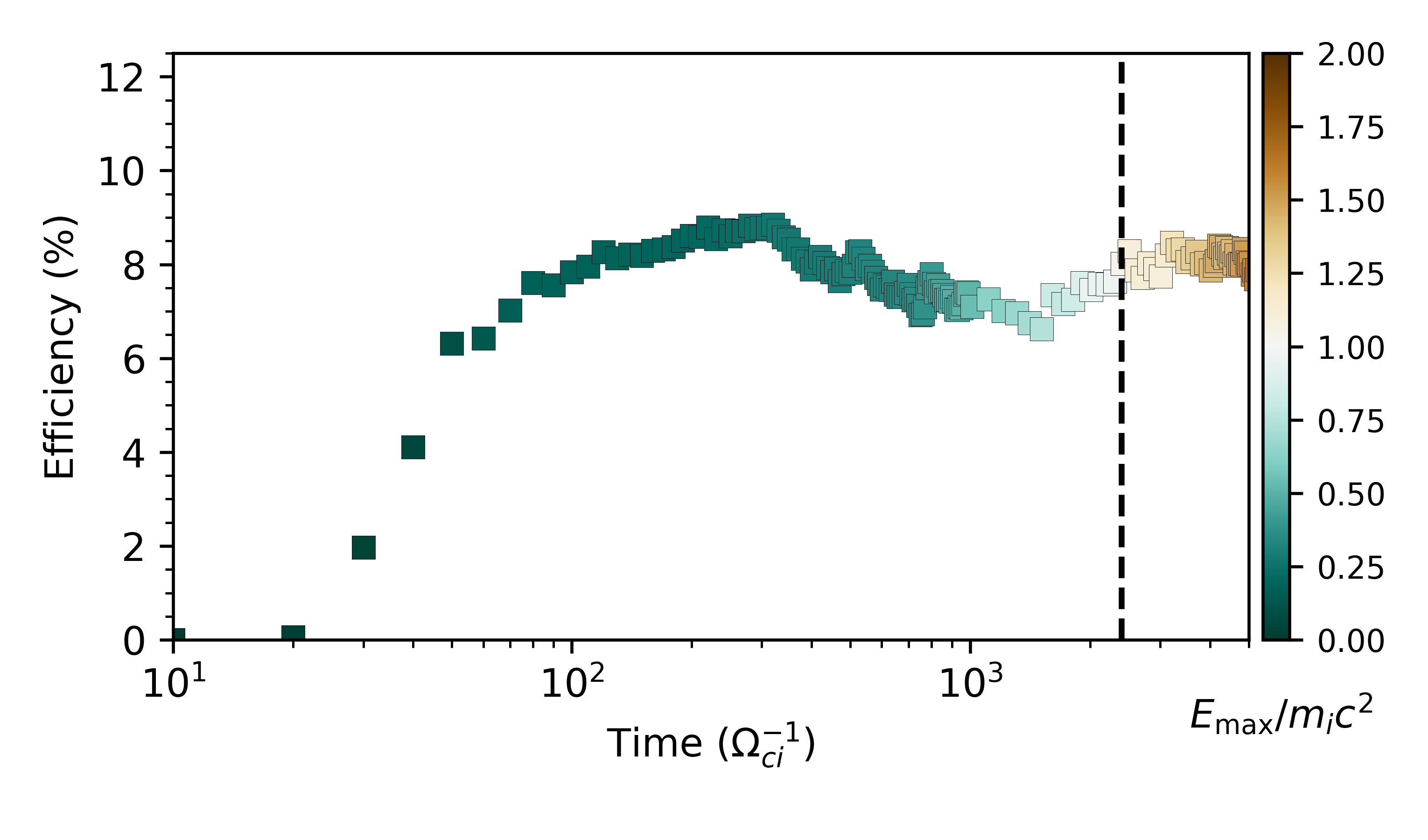}
\caption{CR acceleration efficiency (fraction of energy in particles with $E > 10 E_{sh}$) as a function of time for Run A. The black dashed vertical line corresponds to when the highest energy particles become relativistic. The color of each point corresponds to $\Emax/m_ic^2$ as shown in Figure~\ref{fig:Emax}.
}
\label{fig:acc}
\end{figure}

Figure~\ref{fig:acc} shows that an acceleration efficiency on the order of $10\%$ is reached within the first hundred inverse cyclotron times, and then remains nearly constant throughout the entire simulation, consistent with what was seen in the non-relativistic case \citep{caprioli+14a}. The vertical black dashed line denotes when $\Emax \sim m_ic^2$, the color corresponds to $\Emax / m_ic^2$ as shown in Figure~\ref{fig:Emax}.

From this it is clear that $\ecr$ is unaffected as the CR population transitions from non-relativistic to relativistic energies, and that the canonical value of $\sim 10\%$ quoted by \citep{caprioli+14a} should be considered the asymptotic one. 
In this respect, it is worth stressing that in the non-relativistic regime the efficiency $\ecr\propto E^2 f(E)\propto E^{1/2}$ is typically dominated by the highest-energy CRs, while in the relativistic regime there is about the same energy density per decade. 
Since $\ecr$ saturates well before CRs become trans-relativistic, it is necessary for the shock to ``be aware'' of the efficient CR acceleration;
such a CR feedback will be discussed in greater detail in forthcoming works, but here we mention that the pressure in the CR precursor affects the dynamics shock front, which reacts by injecting fewer particles into DSA.

Until this moment we have not discussed oblique or perpendicular shocks.
This is because it has previously been found in classical hybrid simulations that shocks with $\vartheta_{Bn}\gtrsim 50^{\circ}$, thermal ions are not energized enough to initiate the DSA process \citep{caprioli+14a, caprioli+15}. 
Note that, if the injection issue is overcome, for instance when pre-energized CR seeds are present \citep{caprioli+18}, or the presence of external plasma turbulence, acceleration at oblique shocks proceeds unhindered, even more rapidly than at quasi-parallel shocks \citep[e.g.,][]{jokipii87, giacalone05}

\begin{figure}
\includegraphics[width=.49\textwidth,clip=true,trim= 0 0 0 0]{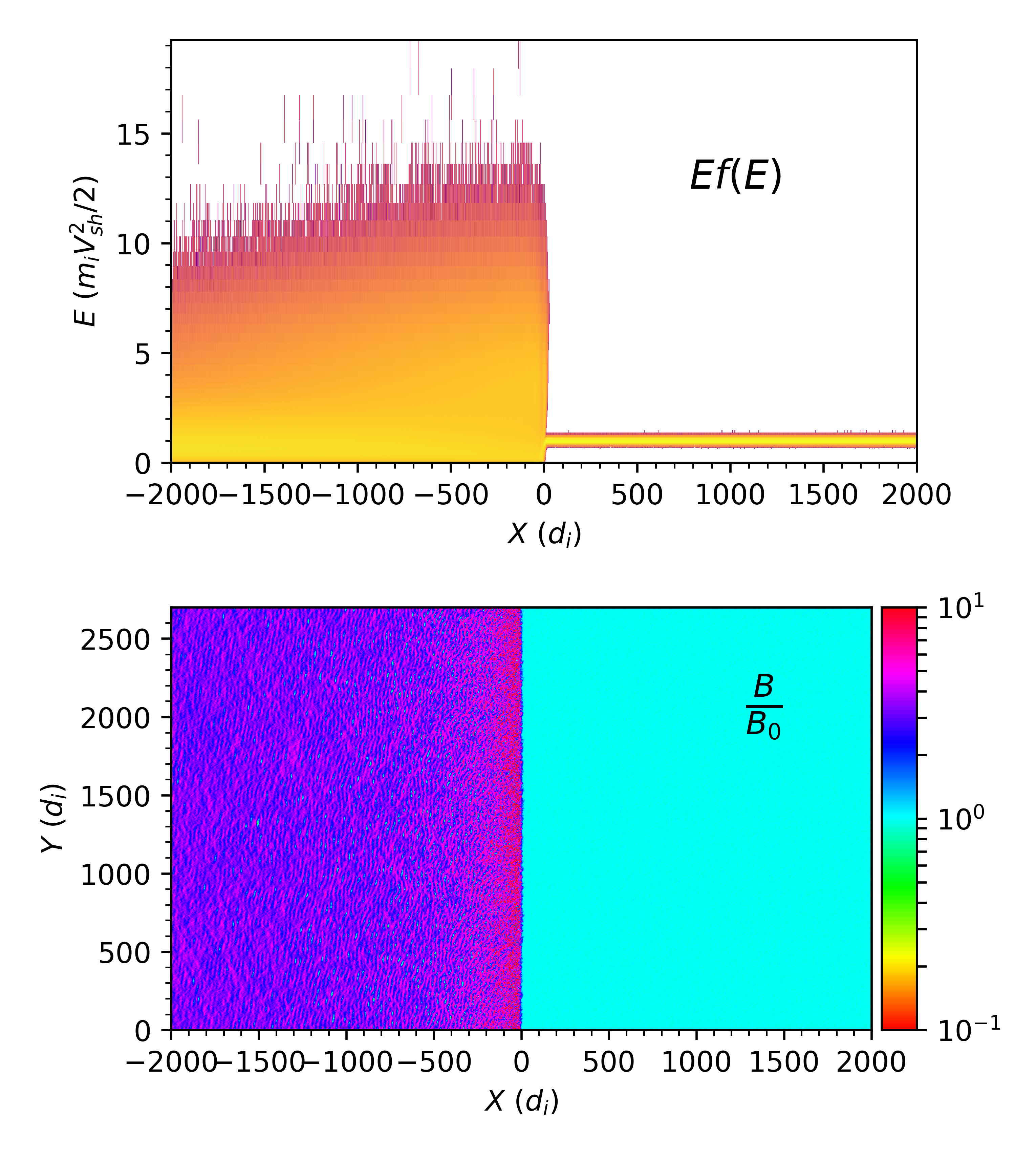}
\caption{Quantities from the quasi-perpendicular shock simulation described by Run C in Table~\ref{tab:shocks}.
Top panel: Energy spectrum at each position $x$;
Bottom panel: 2D plot of the magnitude of the magnetic field.
There is no evidence of DSA and of magnetic field amplification upstream.}
\label{fig:no_perp_inj}
\end{figure}

Recently, PIC-MHD simulations of very oblique shocks ($\vartheta_{Bn}\gtrsim 70^{\circ}$) have suggested that thermal particle injection and DSA will eventually occur for simulations run long enough \cite{vanmarle+18}. 
We have tested this claim with the full-hybrid \dHybridR{} code and did not recover such a result.
Figure~\ref{fig:no_perp_inj} shows a simulation perform with the same initial parameters  as the quasi-perpendicular, $M=30$  simulation discussed in \cite{vanmarle+18}. 
The simulation is $[L_x;L_\perp]=[10^4; 2.7\times 10^3]d_{i0}$ in size with two cells per skin depth in each direction and was run for a comparable amount of time ($600 \Omega_{ci}^{-1}$). 
Using 4 particles per cell, the \dHybridR{} simulation has approximately $4/3 \times M\times 16\times  2700\simeq 1.72\times 10^6$ macro-particles impinging on the shock per unit cyclotron time, where the factor of $r/(r-1)\simeq 4/3$ comes from the conversion of the upstream flow speed from the simulation to the shock frame.
For the canonical 1\% injection efficiency \citep{caprioli+15}, in our simulation $\sim
1.7\times 10^4$ CR particles are produced per unit time, which returns a  statistics comparable with the $\sim 10^4$  rate used by \cite{vanmarle+18}.
The top panel shows the energy density distribution as a function of $x$, in which energy is normalized to the shock energy. 
Downstream of the shock ($x<0$), ions are heated up to supra-thermal energies ($E \lesssim 10 E_{sh}$), but there is no DSA tail, and no energetic particles upstream ($x>0$). 
The bottom panel of Figure~\ref{fig:no_perp_inj} shows a 2D plot of the magnitude of the magnetic field, which reveals the canonical downstream compression, with additional some small-scale deviations (which we discuss below); the upstream magnetic field, instead, is unperturbed.
These results stress how a self-consistent model for ion injection can only be provided by full-hybrid simulations.

\section{3D Simulations}\label{sec:3D}
\begin{figure*}
\includegraphics[width=\textwidth]{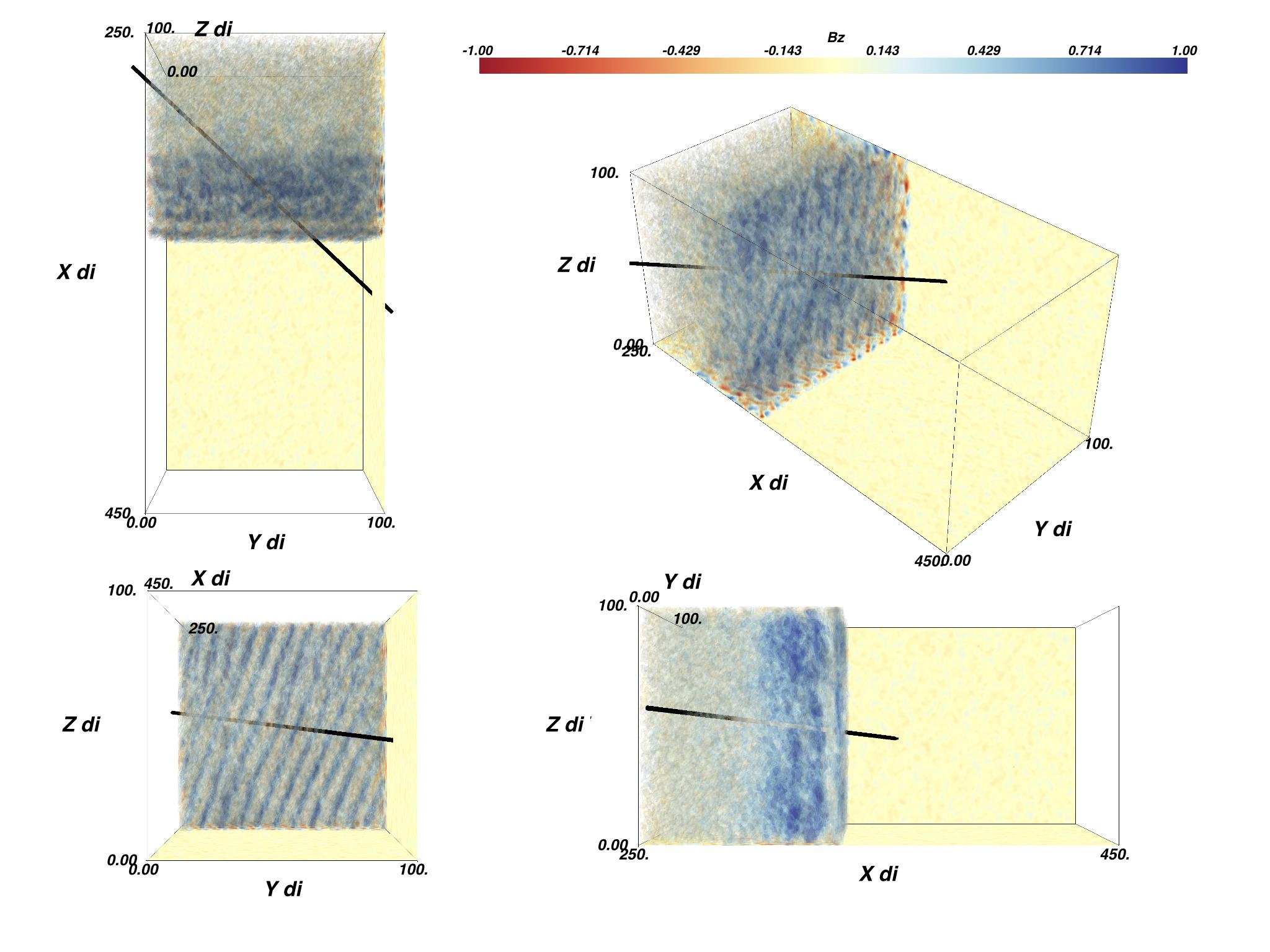}
\caption{Orthographic projection of $B_z$ (self-generated component normal to the upstream flow and mean magnetic field) around the shock from Run 3D. 
Four views of the 3D structure, with the following views from bottom right in clockwise order: viewing along $+{\bf \hat{y}}$, along $-{\bf \hat{x}}$, along $-{\bf \hat{z}}$ and an isometric view along $(-{\bf \hat{x}}+{\bf \hat{y}}-{\bf \hat{z}})/\sqrt{3}$. Slices of $B_z$ in the $x,y$ and $x,z$ are plotted along the edge of the plotting domain. The black line represents the trajectory of the 1D cut shown in Figure~\ref{fig:3D_cut}.}\label{fig:3D}
\end{figure*}

Finally, we present a quasi-perpendicular 3D shock simulation with a smaller Mach number ($M = 5$), identified as Run 3D in Table~\ref{tab:shocks}. 
The conditions in this simulation are quite similar to typical heliospheric shocks, such as the  Earth's bow shock, which is formed by the supersonic/super-Alfv\'enic solar wind, traveling at speed $\gtrsim 100$km s$^{-1}$ and  impinging on the Earth's magnetosphere \citep{sheeley+85, cane+03}.
For typical solar wind conditions, the ion temperature is of the order of $10$eV and thermal and magnetic pressure are comparable to each other, which corresponds to $M\approx 5-10$ \citep[e.g.,][]{schwartz+88,wilsoniii+18p};
also interplanetary shocks triggered by coronal mass ejections typically span the same range of Mach numbers  \citep[e.g.,][]{wilson+19}.
\dHybridR{} is well suited to study low-Mach-number heliospheric shocks because in this systems ions can be accelerated to trans-relativistic energies, and because the relevant sizes and scales can be modeled to scale at a reasonable computational cost.
In Fig~\ref{fig:3D} we present an orthographic projection of $B_z$, where $z$ is the direction normal to the upstream flow and the mean upstream magnetic field;
therefore, $B_z$ is the self-generated component of the magnetic field. 
Upstream of the shock there are no indications of magnetic field amplification, in agreement with the 2D simulation. 
However, downstream some magnetic structures can be observed: there is a clear rippling of the magnetic field along the shock interface, which is produced by shock reformation, consistent with what has been previously found in observations \citep{johlander+16b, johlander+18} and simulations \citep{lb03,caprioli+15, burgess+16}. 

\begin{figure}
\includegraphics[width=.49\textwidth,clip=true,trim= 0 0 0 0]{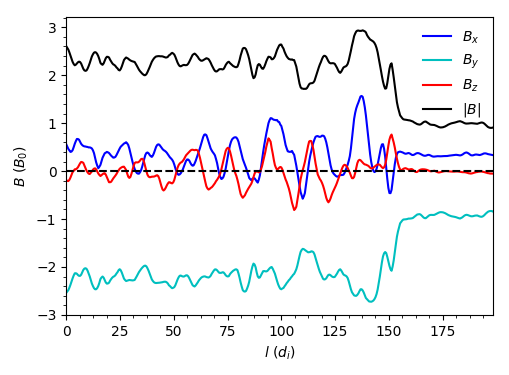}
\caption{The 3 components and magnitude of the magnetic field along an oblique trajectory across the shock (from Run 3D Table~\ref{tab:shocks}). The trajectory is denoted by the black line in 3D projection view in Figure~\ref{fig:3D}.}\label{fig:3D_cut}
\end{figure}

The black line in Figure~\ref{fig:3D} represents the trajectory of a synthetic probe through the simulation box, mimicking in-situ spacecraft observations, and Figure~\ref{fig:3D_cut} shows the magnetic field measured by such a probe. 
The trajectory is diagonal through the shock interface, with only a small component normal to the upstream magnetic field ($0.681{\bf \hat{x}} + 0.727{\bf \hat{y}} - 0.091{\bf \hat{z}}$ intersecting a point in the middle of the $y-z$ plane at $x = 312.5 d_i$). 
From this cut, the periodic structure of the ripples can be clearly seen;
considering that the direction of propagation is primarily in the $y$ direction, the wave number can be estimated to be on the order of $k \Omega_{ci}/v_{\rm sh} \sim k r_g \sim 1$, where $r_g$ is the gyroradius of the downstream population.
This is a great example of how \dHybridR{} simulations can be directly compared with {\it in-situ} measured heliospheric plasma phenomena. 

\section{Conclusion}
In this work we presented the first results from \dHybridR{}, a hybrid plasma simulation code that includes relativistic ion dynamics.
We detail how relativistic ion motion is included in the code and how for specific systems of interest, the assumptions required for hybrid simulations are not violated.
This novel simulation software can be used to help understand, from first principles, numerous different open problems involving space and astrophysical plasmas.
The code is well suited to study many astrophysical systems where a high energy, low density CR population interacts with a non-relativistic thermal background population.

To verify that \dHybridR{} can correctly model physical systems of interest, we simulated CR-driven non-resonant and resonant streaming instabilities.
In both test cases, the location in $k$ space and the value of the maximum growth rate found in simulations agreed remarkably well with the linear prediction.
Then, we moved to use  \dHybridR{} to model strongly non-linear problems such as DSA at non-relativistic collisionless shocks, similar to those found in the heliosphere, in SN remnants, and in galaxy clusters. 
In particular, we presented simulations with parameters relevant to fast SN shocks (radio SNe, Figure \ref{fig:overview}) as well as heliospheric shocks such as the Earth's bow shock (Figure \ref{fig:3D}).

We performed unprecedentedly-long simulations of parallel shocks in which ions achieve Lorentz factors as large as $\gamma \gtrsim 20$, attesting for the first time in full hybrid simulations that  DSA produces a power-law tail in momentum across the trans-relativistic regime, which implies an energy distribution that follows a broken power law that steepens by 0.5 in slope.
When CRs become relativistic, the increase of the maximum particle energy is still linear in time, but with a rate reduced by a factor of $\sim 2$;
such a reduction is a consequence of the saturation of the velocity of escaping particles to $c$.

The acceleration efficiency (i.e., the fraction of the shock energy channelled into non-thermal particles with energy $E\gtrsim 10 E_{sh}$) was found to reach about $10\%$ within tens of cyclotron times and remain nearly constant as the high energy population transitions into the relativistic regime. 
These results are directly applicable to fast radio SNe, where we predict GeV/TeV CRs to be produced within seconds/days. 
With the current sensitivity of $\gamma$-ray and neutrino telescopes, such a delay could be measured for a Galactic SN.

Finally, we presented a 3D simulation produced with \dHybridR{} with conditions comparable to the Earth's bow shock with a quasi-perpendicular configuration. 
We showed that \dHybridR{}  reproduces both qualitatively and quantitatively the shock rippling that has been found with {\it in-situ} satellite observations \citep{johlander+18}. 

In summary, this work presents, to the authors' knowledge, the first hybrid simulations to include relativistic ion dynamics, which is a critical tool for studying the inherently multi-scale nature of CR/thermal ion interplay in space and astrophysical plasmas.

\acknowledgments
We would like to thank Luis Gargat\'e for providing the original version of {\tt dHybrid}. 
This research was partially supported by NASA (grant NNX17AG30G, 80NSSC18K1218, and 80NSSC18K1726) and NSF (grants AST-1714658 and AST-1909778).
Simulations were performed on computational resources provided by the University of Chicago Research Computing Center, the NASA High-End Computing Program through the NASA Advanced Supercomputing Division at Ames Research Center, and XSEDE TACC (TG-AST180008).

\bibliography{Total}
\bibliographystyle{aasjournal}

\end{document}